# Cube Handling In Backgammon Money Games Under a Jump Model

Mark Higgins

April 2012

## Abstract

A variation on Janowski's cubeful equity model is proposed for cube handling in backgammon money games. Instead of approximating the cubeful take point as an interpolation between the dead and live cube limits, a new model is developed where the cubeless probability of win evolves through a series of random jumps instead of continuous diffusion. Each jump is drawn from a distribution with zero mean and an expected absolute jump size called the "jump volatility" that can be a function of game state but is assumed to be small compared to the market window.

Closed form approximations for cubeful equities and cube decision points are developed as a function of local and remote jump volatility. The local jump volatility can be calculated for specific game states, leading to crisper doubling decisions.

## 1. A new model for game evolution

*Background*

Rick Janowski(1) built on work by Keeler and Spencer(2) who modeled a backgammon game in an abstract but useful way: the probability of winning the game was treated as a Brownian motion, starting at 50%, and eventually diffusing continuously either to 0% (a loss), 100% (a win), or to a point where a player offered the doubling cube to their opponent.

In this "live cube limit" it is always optimal to hold onto the cube as long as possible, and the correct time to double is right at the opponent's take/pass threshold (the "cash point"): if the player doubles just below the threshold he gets the maximal post-double value. And because the probability of winning diffuses continuously, the player never has to worry about getting a great roll and having the game jump to a state where he has "lost his market" because on a double his opponent has a clear pass. This makes the live cube limit an upper bound to the true cubeful equity.

Janowski corrected for this bias by noting that the "dead cube limit", where no value is assigned to holding the cube, is a lower bound. He then modeled the true cubeful equity as a linear interpolation between the dead cube equity and the live cube equity for each of the three cube states (centered, owned by the player, and unavailable - owned by the opponent).

That interpolation parameter, called the cube life index, is designed to perform a similar role to game state jumps in the model presented here: accounting for the true value of owning the cube.

He noted that really the cube life index $x$ should be a function of game state, and depends on the distance (in game state) to the optimal doubling point, width of the doubling window, and volatility of the position; and that it should be different for each player. In practice most backgammon bots assume a single constant cube life index, or define a heuristic form for $x$ as a function of a small number of broad game states.

Janowski's other innovation was in a simple representation of the game state that incorporated gammons and backgammons. Previous work had restricted itself to games with no gammons or backgammons, and proxied the game state with the cubeless probability of winning. Janowski continued to use the win probability as the game state proxy, but assumed that the expected number of points on a win, and the expected number of points on a loss, remained constant as the win probability moved.

With expressions for the three cubeful equities as a function of the cubeless probability of the player winning, $P$, the cube decision points can be determined. The take point $TP$, where the player is on the margin between taking the cube and passing on it when the opponent offers, is

$$E_O(TP) = -\frac{1}{2}$$

where $E_O(P)$ is the cubeful equity when the player owns the cube, normalized to the current cube level. If $E_O(P)$ is less than $-{}^1\!/_2$ then the player should pass when offered the cube and lose a value equal to the current cube $C_V$; otherwise she should take and end up with equity $2C_VE_O(P)$.

Similarly the cash point $CP$ is where the opponent is on the threshold of take vs pass when the player owns the cube:

$$E_U(CP) = \frac{1}{2}$$

where $E_U(P)$ is the cubeful equity when the opponent holds the cube, normalized by the cube value.

The redouble point $RD$ is where a player who owns the cube will choose to redouble his opponent:

$$2E_U(RD) = E_O(RD)$$

and the initial double point $ID$ is where a player will double when the cube is centered (Jacoby rule aside, this will be less than $RD$):

$$2E_U(ID) = E_C(ID)$$

where $E_C(P)$ is the cubeful equity when the cube is centered.

There are also too-good points when the cubeful equity at the current cube level is equal to 1 – past that it no longer makes sense for a player with access to the cube to double. There is one too-good point for the centered cube and one for the player-owned cube; the centered-cube too-good point is larger than the player-owned cube too-good point. Both are at $P$=100% if the player has no chance of a gammon.

*The jump model*

This paper addresses the same problem of optimal doubling and take/pass strategy by developing a different model: one where the probability of win takes discrete jumps from turn to turn instead of continuously diffusing. Each jump is modeled as an independent sample from a jump distribution, with mean zero and an expected absolute jump size called the **jump volatility**. Jump volatility is allowed to be a function of game state. Note that jump volatility is not the standard deviation of the jump distribution; it is the expected value of the absolute value of the jump. That is the most important feature of the jump distribution for cubeful equity and cube decision point calculations.

Moving from continuous diffusion to jumps addresses the core weakness of the live cube limit: it ignores the possibility of jumping across the take/pass threshold and losing equity to the pass.

A jump in probability in this model corresponds to the change in the cubeless probability of win, starting from when the player holds the dice ready to throw (or make a cube decision), ending when the player receives the dice again after the opponent's move. That is, the change in probability from cube decision to cube decision.

I assume that jump volatility is a function of game state and introduce two different jump volatilities needed to calculate cube actions: "local" jump volatility and "remote" jump volatility.

Local jump volatility is the expected absolute jump in cubeless win probability at the instantaneous game state. Remote jump volatility is the expected jump volatility if the game turns around and moves from the player's advantage to the opponent's, and the opponent has an opportunity to redouble.

Appendix 1 derives the take and cash points for a linear approximation to the jump model:

$$TP = \frac{\left(L - \frac{1}{2}\right)(L+1)}{\left(W + L + \frac{1}{2}\right)\left(L + 1 - \frac{\alpha_r}{4}\frac{\left(W+L+\frac{1}{2}\right)}{\left(W-\frac{1}{2}\right)}\right)}$$

$$CP = \frac{L+1}{W+L+\frac{1}{2}} - \alpha_r \frac{W - \frac{1}{2}}{2(2LW + 2L - W - 1) - \alpha_r\left(W + L + \frac{1}{2}\right)}$$

$W$ represents the expected points won conditioned on a win, as in Janowski's model; $L$ represents the expected points lost conditioned on a loss; $\alpha_r$ is the remote jump volatility.

Appendix 1 also includes expressions for the cubeful equities in the three cube states (centered-cube, player-owned cube, and unavailable cube) in the linear approximation. These can depend on the local volatility $\alpha_l$ which factors into initial double and redouble point calculations.

I also develop a more accurate nonlinear approximation that improves on the linear approximation for high jump volatilities. In most cases, however, the linear approximation is sufficient. Appendix 2 validates the approximation against an exact numerical solution of the cubeful equities.

# 2. Estimating the jump volatility

*Estimating remote jump volatility*

Estimating jump volatility "over the board" as a human is challenging, but fortunately we have backgammon bots that play very credible games. I trained an intermediate-strength backgammon bot that uses neural networks to estimate the probability of win, as well as the probability of gammon and backgammon outcomes[5]. I let the bot play money games against itself using either the Janowski model or the jump model to make cube decisions. I am assuming that the statistics that result from these calculations are sufficiently accurate even though the bot is not a perfect player.

First I investigated self-play of the bot, the player following the nonlinear approximation of the jump model for different constant jump volatilities, the opponent following the Janowski model with a fixed cube life index of 0.70 (the optimal cube life index I get from self-play of my bot[5]).

The average points per game scored by the jump model player are displayed below. A polynomial fit through the results indicates an optimal constant jump volatility of 9.1%. This is the most accurate measurement of remote jump volatility I can make since self-play picks out exactly the statistics that matter most to the doubling strategy.

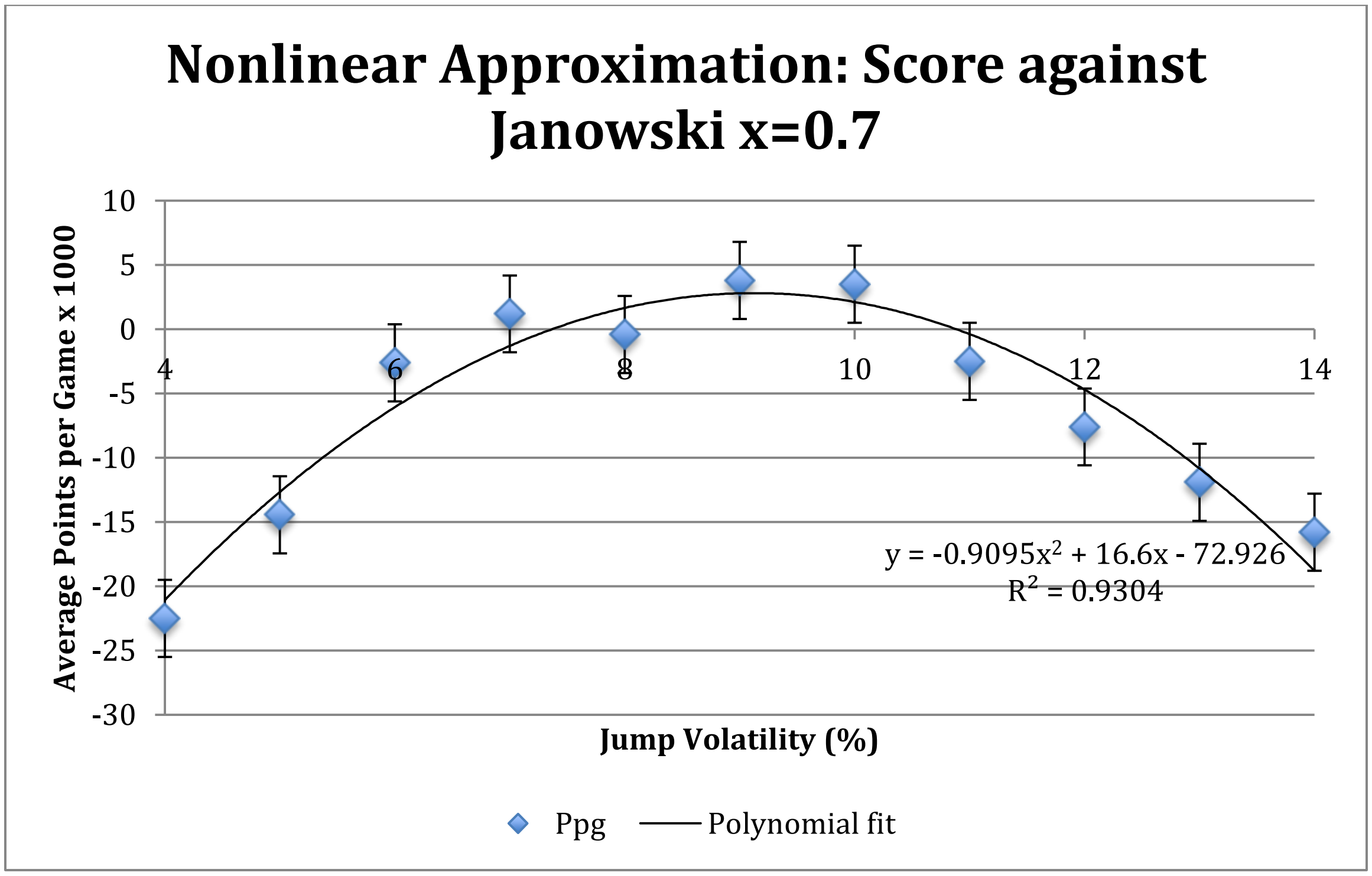

Figure 1: average milli-points per game won in money games against an opponent using a Janowski model strategy with fixed $x$=0.7, by a player using the nonlinear approximation to the jump model with different constant values of the jump volatility. All average points per game numbers had a standard error of 3 milli-ppg.

I also implemented the linear approximation to the jump model equity and played that against the same Janowski model opponent. The chart below shows the average points per game scored at different constant jump volatility levels.

A polynomial fit through the results suggests an optimal constant jump volatility of 11.3%. While the performance of the optimal player is roughly the same as with the nonlinear approximation, the optimal jump volatility here is somewhat higher, presumably correcting for the imperfections of the linear approximation.

The optimal scores for the linear and nonlinear approximations were very close, suggesting that the linear approximation is sufficient. However, jump volatilities must be scaled up from statistical estimates by a factor of 11.3/9.1=1.24 before they are passed into the linear approximation, to account for inaccuracies in the approximation. Future work could investigate this bias further.

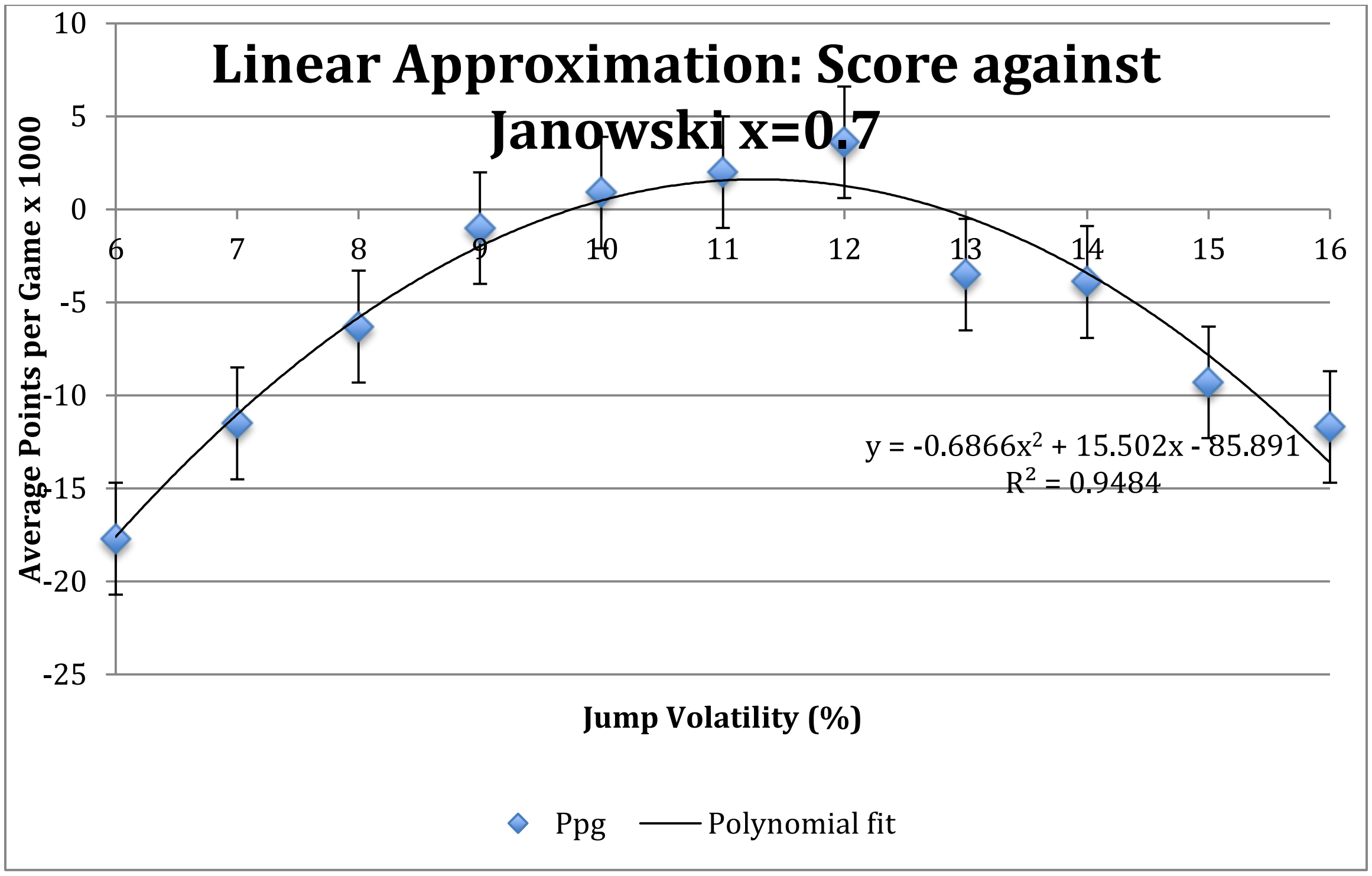

Figure 2: average milli-points per game won in money games against an opponent using a Janowski model strategy with fixed *x*=0.7, by a player using the linear approximation to the jump model with different constant values of the jump volatility. All average points per game numbers had a standard error of 3 milli-ppg.

Another estimate of remote jump volatility can be made by explicitly calculating the expected value of the absolute value of probability jumps from turn to turn.

When calculating take and cash points we need an estimate of the remote jump volatility conditioned on first landing in a game state where one of the players would take, then subsequently landing on a state where the same player would redouble. I ran 100,000 cubeless money games of my backgammon bot against itself. I filtered down the game states to ones where the probability of win passed through either 20-35% or 65-80%; then passed back through the opposite window. If both conditions were met I calculated the change in probability of win over two plies and added that change to the set. Finally I calculated the standard deviation of probability changes across the set to get the remote jump volatility.

The remote jump volatility (the expected absolute value of the cubeless win probability jump) for this set was 9.4%, averaged over 81k game states that passed the filter.

This is quite close to the optimal jump volatility of 9.1% found from bot self-play using the nonlinear approximation to the jump model, suggesting again that if jump volatility is estimated statistically, it should be scaled up by a factor of 1.24 before using in the linear approximation formulas.

This statistical estimate is averaged across all game states and may not be flexible enough: as with the Janowski cube life index we may want to look at using different remote jump volatility values for different game states.

For example if we split that set into states where, at the first of the two windows, the game state is a race, the remote jump volatility if the game turns is 14.8%. For contact games, the remote jump volatility is 9.3% (most cases of starting in one region and migrating to the other were initially contact games).

*Estimating local jump volatility*

The real benefit of this model is the ability to estimate local jump volatility for specific game states. This lets a player make a more accurate doubling decision than if he used a constant jump volatility, or even a jump volatility that is a function of a small number of heuristic game states.

The local jump volatility can be calculated by a bot by calculating the standard deviation over the 21x21=441 possible rolls in the next two rolls of the dice (one for the player, one for the opponent):

$$\sigma_J = \sqrt{\sum_{i=1}^{N} w_i (P_i - P_a)^2}$$

and

$$P_a = \sum_{i=1}^{N} w_i P_i$$

where $N$=441 (one for each of the pair of possible rolls), $P_i$ is the cubeless probability of win in the $i^{\text{th}}$ state after two rolls (and the bot makes the optimal move for each roll), $P_a$ is the average cubeless probability of win across the end states, and $w_i$ is the weight assigned to the pair of rolls ($1/36^2$ if both rolls are doubles; $1/18^2$ if both rolls are mixed; and 1/(18*36) if one roll is a double and the other mixed). Note that I do not assume that the average probability equals the probability in the starting state. This would be true for a perfect bot evaluation function, and will be close for sufficiently strong bots, but we may as well remove that bias since we can do it easily.

This two-step lookahead is often quite an expensive calculation for a backgammon bot. One possibility, suggested by Øystein Johansen, is to approximate the jump volatility as a function of game state using a neural network trained on precalculated local jump volatility for a range of benchmark board layouts. This is not required, however, if the backgammon bot is already doing a two-step lookahead as part of its normal board evaluation, in which case the required probabilities are already available.

As noted earlier, if these statistically-estimated jump volatilities are applied to the linear approximation, they should be scaled up by a factor of 1.24 to account for the inaccuracy of the linear approximation.

# 3. Connection to Janowski's model

The linear approximation to the jump model has an interesting connection to Janowski's "refined general model" which allows for two cube life indexes, one for each player. The take point in this model is a function only of $x_1$ and the cash point only of $x_2$.

If I assume a constant jump volatility $\alpha$ and calculate the implied $x_1$ and $x_2$ from the jump model take and cash points I get

$$x_1 = 1 - \frac{\alpha\left(W + L + \frac{1}{2}\right)^2}{2(L+1)(W-\frac{1}{2})}$$

$$x_2 = 1 - \frac{\alpha\left(W + L + \frac{1}{2}\right)^2}{2(W+1)(L-\frac{1}{2})}$$

In a symmetric game state where $W$=$L$ the implied Janowski cube life indexes are equal. But in general game states they are different.

The table below shows the implied cube life index values for different values of $W$ and $L$ and $\alpha$=10%, a typical value. Each value in the table shows the pair of cube life index values as $x_1/x_2$:

| | ***W*=1** | **1.25** | **1.5** | **1.75** | **2** |
|---|---|---|---|---|---|
| ***L*=1** | 0.69/0.69 | 0.75/0.66 | 0.78/0.64 | 0.79/0.62 | 0.80/0.59 |
| **1.25** | 0.66/0.75 | 0.73/0.73 | 0.77/0.72 | 0.78/0.70 | 0.79/0.69 |
| **1.5** | 0.64/0.78 | 0.72/0.77 | 0.76/0.76 | 0.78/0.74 | 0.79/0.73 |
| **1.75** | 0.62/0.79 | 0.70/0.78 | 0.74/0.78 | 0.77/0.77 | 0.78/0.76 |
| **2** | 0.59/0.80 | 0.69/0.79 | 0.73/0.79 | 0.76/0.78 | 0.78/0.78 |

Also interesting is the variation of implied cube life index with $\alpha$. The next table shows its variation for $W$=$L$ where the two implied cube life indexes are equal:

| | **$W$=$L$=1** | **1.25** | **1.5** | **1.75** | **2** |
|---|---|---|---|---|---|
| **$\alpha$=0** | 1 | 1 | 1 | 1 | 1 |
| **5%** | 0.84 | 0.87 | 0.88 | 0.88 | 0.89 |
| **10%** | 0.69 | 0.73 | 0.76 | 0.77 | 0.78 |
| **15%** | 0.53 | 0.60 | 0.63 | 0.65 | 0.66 |
| **20%** | 0.38 | 0.47 | 0.51 | 0.53 | 0.55 |

Using the linear approximation self-play estimate of 11.3% for $\alpha$ and $W$=$L$=1.27, representing the average win and loss points at the start of the game, I get an implied $x$=0.70, equal to the optimal cube life index of 0.70 I found through Janowski model bot self-play. This is also consistent with other estimates of cube life index by Johansen[3] and the GNU Backgammon team[4].

Another interesting connection is that if we choose $x_1$ and $x_2$ to equal their implied values, the other Janowski cube decision points match those of the linear approximation to the jump model.

That fact implies that this model gives a method for choosing the Janowski cube life indexes in different game states.

However, that conclusion assumes we are using a constant jump volatility and not taking advantage of the jump model's ability to estimate the local jump volatility accurately. When the jump model is generalized to include a local volatility the direct connection to Janowski's model is broken. The take and cash points depend only on remote jump volatility so the implied Janowski cube life indexes are not a function of local volatility. However, the jump model's initial double and redouble points are a function of local jump volatility.

The jump model also includes a more accurate nonlinear approximation that gives (somewhat) different and more accurate cube decision points than the linear approximation. There is no clear connection to Janowski's model in this case.

# 4. Conclusions

I developed a new method of determining cubeful equity in backgammon money games that replaces Keeler and Spencer's continuous diffusion of cubeless win probability with a process that evolves through a series of random jumps. Under the approximation of small but realistic jump volatility, cubeful equity in the three states (centered, player-owned, and unavailable) have closed form expressions in terms of expected win points, expected loss points, and the remote and local jump volatilities.

Two approximations were developed: a linear approximation and a more complex but more accurate nonlinear approximation. In most cases the linear approximation suffices, but for large jump volatilities the nonlinear approximation is noticeably better. Both were validated against an exact numerical solution.

Bot self-play was used to estimate the optimal constant jump volatility in the linear approximation: 11.3%. This is equivalent to a Janowski cube life index in the range 0.65-0.75, depending on $W$ and $L$. The optimal jump volatility from self-play for the nonlinear approximation was 9.1%. This play-based estimate is comparable to a separate statistical estimate of jump volatility from turn-to-turn jumps in cubeless win probability of 9.4%.

If a constant jump volatility is used in the linear approximation, the model can be used to imply state-dependent Janowski cube life indexes in his refined general model.

# Acknowledgements

I would like to thank Rick Janowski for reviewing this paper and providing critical insights into his model, ideas on local and remote jump volatility, and generally how cubeful equity calculations should work for different cube states. Also Øystein Johansen for ideas on ways to estimate local jump volatility.

# Appendix 1: the jump model

The live cube model above is not as realistic as we would like because probability of win does not really diffuse. In reality it steps from one roll to the next, and stepping across the cash point when you own the cube loses you equity that you would be able to capture in the live cube model.

*Cubeful equity evolution*

Over a full round – going from the player holding the dice before throwing, past the opponent owning the dice, back to the player holding the dice again – the cubeless probability of win will jump due to the player's roll and the opponent's roll. The jump in win probability causes a jump in equity: the "cubeful" equity, which means it accounts for cube actions. This is opposed to the "cubeless" equity that ignores the possibility of cube actions.

The current equity is just the expected value of the equity after the full round has passed:

$$E(P) = \int_{J=-\infty}^{\infty} f_P(J)\, E(P+J)\, dJ$$

where $E(P)$ is the cubeful equity at the current game state (proxied by the cubeless probability of win $P$) and $f_P(J)$ is the distribution of the jump in $P$. It has a $P$ subscript to denote that the jump distribution can be a function of game state. The mean of the jump distribution must be zero: the expected value of the cubeless win probability after the next turn must equal its current value.

In reality there are only 21x21=441 possible states that the game can move to after the two rolls, but we approximate that discrete distribution with a continuous one. And of course the jump distribution does not really extend across the whole real line, since the probability $P$ cannot jump below 0% or above 100%; but we will be assuming narrow jump distributions later and that approximation will not be so severe.

Following Janowski I subsume the rest of the game state into two variables $W$ and $L$. $W$ corresponds to the expected number of points the player wins conditioned on a win; $L$ corresponds to the expected number of points the player loses conditioned on a loss (a positive number). I assume $W$ and $L$ stay constant as the probability of win $P$ moves. These values are calculated as

$$W = \frac{P + P_{GW} + P_{BW}}{P}$$
$$L = \frac{1 - P + P_{GL} + P_{BL}}{1 - P}$$

where $P$ is the cubeless probability of any kind of win, $P_{GW}$ and $P_{GL}$ are the cubeless probabilities of winning and losing any kind of gammon, and $P_{BW}$ and $P_{BL}$ are the cubeless probabilities of winning and losing a backgammon. All these values are typically available to backgammon bots from neural net calculations.

Now we need to consider the cubeful equity in different cube states. I introduce three different cubeful equities: $E_C(P)$, the equity for a centered cube; $E_O(P)$, the equity when the player owns the cube; and $E_U(P)$, the equity when the cube is unavailable because the opponent owns it. All these equities are normalized to the current cube value, so run between $-L$ at $P$=0% and $+W$ at $P$=100%.

If the cube is centered or the player owns the cube, he has the opportunity to double before his dice throw. I assume this happens optimally: he always decides to double if it is to his advantage from an equity perspective. Similarly I assume the opponent makes optimal decisions when she has the option of doubling.

Those assumptions lead to relationships between the different cubeful equities:

$$E_O(P) = \int_{J=-\infty}^{\infty} f_P(J) \max(E_O(P+J), \min(2E_U(P+J), 1))\, dJ$$

$$E_U(P) = \int_{J=-\infty}^{\infty} f_P(J)\min(E_U(P+J), \max(2E_O(P+J), -1))dJ$$

$$E_C(P) = \int_{J=-\infty}^{\infty} f_P(J)\min\big(\max(E_C(P+J), \min(2E_U(P+J), 1))\,, \max(2E_O(P+J), -1)\big)dJ$$

Consider the first equation, corresponding to the equity when the player holds the cube. The integrand corresponds to what the equity will be after the jump in probability. At that point the player looks at the equity if he does not double (still $E_O$) and if he doubles (the minimum of the equity where the opponent holds the cube, $E_U$, and the cash amount +1), and chooses the cube action that gives him the maximum of the two. Again, this is assuming the player makes the optimal doubling decision at each point after the jump.

Similarly the second equation, for unavailable cube equity, represents the optimal doubling decision of the opponent.

The third equation represents equity for a centered cube, and reflects the fact that both player and opponent have the opportunity to double.

*The linear approximation*

First I will develop a linear approximation to the solution of these three equations. I start by assuming that jumps are small: more precisely, that the range over which the jump distribution $f_P(J)$ has significant value is small compared with 1.

Consider a value of $P$ that is far from any cube decision point, where there is no danger of the player or opponent doubling. In these cases the integral equations reduce to

$$E(P) = \int_{J=-\infty}^{\infty} f_P(J)\, E(P+J)dJ$$

This applies to all of the three cubeful equities.

A linear solution $E(P) = A + B\,P$ solves this integral equation for any jump distribution (as long as the average jump is zero, which is must be). So we now know that the cubeful equity is approximately linear when the probability $P$ is far enough from a cube decision point that $P$ has no significant probability of jumping to it in one turn.

Now I take a leap: assume that the equity is piecewise linear. For the player-owned cube equity I assume one breakpoint, at the cash point. For the unavailable cube (opponent-owned) equity I assume a single breakpoint at the take point. And for centered-cube equity I assume two breakpoints, at the take and cash points.

This is a fairly dramatic assumption, and I justify it in Appendix 2 where I develop an exact numerical solution to the three integral equations and test the approximation.

The benefit of this piecewise linear approximation is that it lets us calculate simple expressions for the take and cash points, which in turn give us relatively simple expressions for the three cubeful equities.

When jump volatility is zero the jump model reduces to the live cube limit. So the first guess of the solution should be something close to the live cube limit. The approximation I make first is that we can approximate the equities inside the integrals above as the live cube equities. I use that to calculate estimates of equities at the live cube take and cash points, and then approximate the equity as piecewise linear between the known points.

The advantage of starting with the live cube limit is that it is never appropriate to double before the cash point, and it is always too good to double above the cash point; doubles are offered only in an infinitesimally wide region just below the cash point. This means the post-jump equity in the live cube limit is always the current cube equity.

Consider the player-owned cube equity $E_O(P)$ first:

$$E_O(P) \approx \int_{J=-\infty}^{\infty} f_P(J)\, E_{OL}(P+J)\, dJ$$

The player-owned cube equity in the live cube limit, $E_{OL}(P)$, is piecewise linear. It runs from $-L$ at $P$=0 to +1 at the live cube cash point, and then from +1 at the live cube cash point to $+W$ at $P$=1. The live cube cash point is[1]

$$CP_L = \frac{L+1}{W+L+\frac{1}{2}}$$

Consider the equation at $P = CP_L$:

$$E_O(CP_L) \approx \int_{J=-\infty}^{\infty} f_P(J)\, E_{OL}(CP_L + J)\, dJ$$

$$= \int_{J=-\infty}^{0} f_P(J)\big(A_{1L} + B_{1L}(CP_L + J)\big)dJ + \int_{J=0}^{\infty} f_P(J)\big(A_{2L} + B_{2L}(CP_L + J)\big)dJ$$

where $E_{OL}(P) = A_{1L} + B_{1L}P$ for $P$ less than the live cube cash point $CP_L$ and $E_{OL}(P) = A_{2L} + B_{2L}P$ for $P > CP_L$. That is, the piecewise linear pieces of the live cube equity.

By definition we know that $A_{1L} + B_{1L}CP_L = A_{2L} + B_{2L}CP_L = 1$; this is just saying that the live cube equity equals +1 at the cash point. We also know that $A_{1L} = -L$ and $A_{2L} + B_{2L} = W$ to get the correct values at the edges.

I will make another simplifying assumption here: that the distribution of jumps is symmetric for positive and negative jumps. That means

$$\int_{J=-\infty}^{0} f_P(J)dJ = \int_{J=0}^{\infty} f_P(J)dJ = \frac{1}{2}$$

And one more key definition: the jump volatility $\alpha$:

$$\alpha_P = \int_{J=-\infty}^{\infty} |\,J\,| f_P(J)dJ = 2\int_{J=0}^{\infty} J\, f_P(J)dJ = -2\int_{J=-\infty}^{0} J\, f_P(J)dJ$$

The jump volatility comes up naturally in this approximation, which is why I defined it this way, as opposed to the standard deviation of the jump distribution. In practice the only details of the jump distribution that matter are symmetry (for positive and negative jumps) and the jump volatility. Other moments of the jump distribution are largely irrelevant to the results, even when we move to the nonlinear approximation and the exact numerical solution.

There can be significant differences between the jump volatility (the expected absolute value of the jump) and the jump standard deviation. Two examples may help make this clear: the Gaussian distribution and the double-exponential distribution. Under a Gaussian jump distribution the jump volatility is 80% of the standard deviation. With a double-exponential distribution, jump volatility is 71% of the standard deviation. In general, jump distributions with higher kurtosis (fat tails) have smaller ratios of jump volatility to standard deviation.

Also note that the jump volatility can be a function of game state, which this model proxies with the cubeless probability of win $P$.

These choices let us reduce the equation for model equity at the live cube cash point to

$$E_O(CP_L) \approx 1 - \frac{\alpha_r}{4}\frac{\left(W + L + \frac{1}{2}\right)}{\left(W - \frac{1}{2}\right)}$$

Here I use the remote jump volatility $\alpha_r$, as this will be used to calculate take and cash points. The remote jump volatility is appropriate for those as the post-jump equity depends on jump volatility if the game turns and the other player moves to a winning state. That state is far from the current state, so any measure of local jump volatility is not particularly relevant.

We then join this point to $E_O(0) = -L$ as our linear approximation to the exact solution for $E_O(P)$. That is, we will assume that for *P*<*CP* (the model cash point, assumed to be close to the live cube cash point), $E_O(P) = A_{O1} + B_{O1}P$, and we know

$$A_{O1} = -L$$
$$A_{O1} + B_{O1}CP_L = 1 - \frac{\alpha_r}{4}\frac{\left(W + L + \frac{1}{2}\right)}{\left(W - \frac{1}{2}\right)}$$

We can solve for $B_{O1}$:

$$B_{O1} = W + L + \frac{1}{2} - \frac{\alpha_r}{4}\frac{\left(W + L + \frac{1}{2}\right)^2}{\left(W - \frac{1}{2}\right)(L + 1)}$$

The first part of the expression, $W + L + \frac{1}{2}$, is the slope of the live cube equity; the slope of the model equity is slightly smaller.

The "1" in the subscript for $A_{O1}$ and $B_{O1}$ is meant to denote the index of the piece of the piecewise linear function. For player-owned cube equity the first piece runs from *P*=0 to the cash point *CP* (the model cash point, not the live cube cash point, which we have yet to determine). The second piece runs from the cash point to *P*=100%, which we will signify with $A_{O2}$ and $B_{O2}$.

*Calculating the take and cash points in the linear approximation*

This linear approximation is valid for *P*<*CP*, and so it is also valid for *P*=*TP*, the take point. This lets us calculate the take point under this approximation. The definition of the take point is

$$E_O(TP) = -\frac{1}{2}$$

which leads to the expression

$$TP = \frac{\left(L - \frac{1}{2}\right)(L+1)}{\left(W + L + \frac{1}{2}\right)\left(L + 1 - \frac{\alpha_r}{4}\frac{\left(W+L+\frac{1}{2}\right)}{\left(W-\frac{1}{2}\right)}\right)}$$

One can perform a similar operation with the unavailable-cube equity to find the coefficients of the linear function in its second piece (from take point to $P$=100%) and the corresponding cash point:

$$A_{U2} = W - B_{U2}$$

$$B_{U2} = W + L + \frac{1}{2} - \frac{\alpha_r}{4}\frac{\left(W + L + \frac{1}{2}\right)^2}{\left(L - \frac{1}{2}\right)(W+1)}$$

$$CP = \frac{L+1}{W + L + \frac{1}{2}} - \alpha_r \frac{W - \frac{1}{2}}{2(2LW + 2L - W - 1) - \alpha_r\left(W + L + \frac{1}{2}\right)}$$

The cash point has a more complex expression than the take point, but still fairly simple to implement.

Now that we have the take and cash points we can complete the linear approximation for the played-owned and unavailable cube equities by forcing the piecewise linear functions to match at either the cash point (player-owned) or take point (unavailable):

$$A_{O2} = W - B_{O2}$$

$$B_{O2} = \frac{W - (A_{O1} + B_{O1}CP)}{1 - CP}$$

$$A_{U1} = -L$$

$$B_{U1} = \frac{A_{U2} + B_{U2}TP + L}{TP}$$

With that we have the cubeful equity for both player-owned and unavailable cube equities in the model, using a constant volatility equal to the remote jump volatility.

*Adding a local jump volatility*

One advantage of this jump model is that the model parameter, the jump volatility, is a well-defined statistical quantity and can be estimated as a function of local game state. Here I refine the cubeful equity calculations above to incorporate it.

First, note that the take and cash points really are functions only of the remote jump volatility. The take point depends on the played-owned cube equity after the double, which depends on the jump volatility at the cash point. If the game turns and the state moves to the cash point, it is likely in quite a different state than the current one. Similarly for the cash point expression: it depends on jump volatility post-double at the take point, which is in quite a different game state than the current one.

However, the cubeful equity for the current cube level may well depend on the local jump volatility. This is important for doubling decisions.

Consider first the player-owned cube case. Doubling decisions happen near the cash point, so I will assume the appropriate jump volatility to use at the cash point is the local value.

As before we can calculate the equity at the live cube cash point, but this time we will use the local jump volatility $\alpha_l$:

$$E_O(CP_L) \approx 1 - \frac{\alpha_l}{4} \frac{\left(W + L + \frac{1}{2}\right)}{\left(W - \frac{1}{2}\right)}$$

The linear approximation then joins this point up with $E_O(0) = -L$. This gives us the player-owned cube equity for $P<CP$, where the cash point is the same as above, based on the remote jump volatility.

To calculate the redouble point we need to compare the current-cube equity with the equity if the player offers the opponent the cube – the unavailable cube equity. For that we use the expressions above based on the remote jump volatility, as the form depends on jump volatility at the take point which is relatively far from the current game state.

*The centered-cube equity*

The cubeful equity for the centered-cube case depends on jump volatility at both the take and cash points. Again, the take and cash points are as before, dependent on only the remote jump volatility.

The cubeful equity is needed for doubling decisions, and we will consider only the case where the player is in a position to double. The case where the opponent is in a position to double is symmetric.

When the game state is near the cash point, and potentially in initial double territory, the appropriate jump volatility to use at the cash point is the local jump volatility, and the appropriate value at the take point is the remote jump volatility.

Following a similar derivation to the player-owned cube case, but using the live cube equity for the centered cube case instead, I find the centered-cube equities at the live cube take and cash points are:

$$E_C(CP_L) \approx 1 - \frac{\alpha_l}{6} \frac{\left(W + L + \frac{1}{2}\right)(W + 1)}{\left(W - \frac{1}{2}\right)}$$

$$E_C(TP_L) \approx -1 + \frac{\alpha_r}{6} \frac{\left(W + L + \frac{1}{2}\right)(L + 1)}{\left(L - \frac{1}{2}\right)}$$

Note that the equity at the cash point uses the local jump volatility $\alpha_l$ but the equity at the take point uses the remote jump volatility $\alpha_r$.

We can fit a line between these two points to approximate the centered-cube equity for intermediate points, and can use that to calculate the initial double point by comparing it to the unavailable cube equity calculated earlier (using the remote jump volatility).

*The nonlinear approximation*

The linear approximation developed above is often quite close enough, especially given uncertainty in the value of the jump volatility. However we can do better if we want to get a somewhat more accurate estimate of the true equity.

The general approach: instead of using the live cube equity in the integrals, as we did for the linear approximation, use the linear approximation itself. This can be viewed as a second step in an iteration that, if continued, would converge to the exact solution. However, two steps are generally sufficient to get a very good approximation.

Using the linear approximation makes the integrals somewhat more complex. In the live cube model there is never any case where post-jump the game ends in a double-take or a double-pass position; either $P<CP$ and the equity is not good enough to double, or $P>CP$ and the equity makes it too good to double.

That is not true in the jump model's linear approximation, and so those cases will need to be incorporated into the integrals.

Consider the player-owned equity:

$$E_O(P) = \int_{J=-\infty}^{\infty} f_P(J) \max(E_O(P+J), \min(2E_U(P+J), 1)) dJ$$

There are five break points to consider here: $P$=0; the player's redouble point $RD_O$; the cash point $CP$; the player's too-good point $TG_O$; and $P$=100%. The integral equation becomes

$$\begin{aligned} E_O(P) = & \int_{J=-\infty}^{-P} f_P(J)\, E_{O-}(P+J)\, dJ + \int_{J=-P}^{RD_O-P} f_P(J)\, E_{O1}(P+J)\, dJ \\ & + \int_{J=RD_O-P}^{CP-P} f_P(J)\, 2\, E_{U2}(P+J)\, dJ + \int_{J=CP-P}^{TG_O-P} f_P(J)\, dJ \\ & + \int_{J=TG_O-P}^{1-P} f_P(J)\, E_{O2}(P+J)\, dJ + \int_{J=1-P}^{\infty} f_P(J)\, E_{O+}(P+J)\, dJ \end{aligned}$$

Term by term:

- The first corresponds to the part of the integral corresponding to jumps that land at $P$<0. A properly realistic jump distribution would have zero density outside (0,1), but we can get by with simple jump distributions if we fake up equity for $P$<0 (and $P$>1) so that we guarantee to match the known equities at $P$=0 (-$L$) and $P$=1 (+$W$).
- The second term corresponds to jumps that land in 0<$P$< $RD_O$, the player's redouble point. Here the player will not double and the equity we will use is the linear approximation $E_{O1}(P) = A_{O1} + B_{O1}P$.
- The next corresponds to jumps that land in $RD_O$<$P$<$CP$, the cash point. In this region the player will redouble and the appropriate equity is twice the (cube-normalized) linear approximation to the unavailable cube equity.
- The fourth term corresponds to jumps that land in $CP$<$P$< $TG_O$, the player's too-good point. Here the player will double and the opponent will pass, and the player gets the cash amount of +1.
- The fifth term corresponds to jumps that land in $TG_O$<$P$<1, where the player's equity at the current cube level makes it too good to double. We use the linear approximation to the played-owned cubeful equity in the second region.

- The final term is symmetric with the first term, and corresponds to jumps beyond $P$=100%. Again, we fake up the equity here to ensure that at $P$=100% the equity equals the known value of $+W$.

The cube decision points here ($RD_O$, $CP$, and $TG_O$) are the approximate ones that come from the linear approximation of the various equities.

To evaluate those integrals we need more knowledge of the jump distribution than just its symmetry and the jump volatility: we need the full distribution. In principle this means that more details of the distribution than just the jump volatility are relevant to the equity calculation. In practice the difference in equity from using two different distributions, so long as their jump volatilities are the same, is negligible.

We will need two functions of the distribution:

$$F(J) = \int_{x=-\infty}^{J} f(x)dx$$

$$G(J) = \int_{x=-\infty}^{J} x\, f(x)dx$$

The first is the cumulative distribution. The second is similar but weighted by the jump size.

Substituting the linear forms for the equities inside the integrands I get:

$$\begin{aligned}E_O(P) = {} & (A_{O-} + B_{O-}P)F(-P) + B_{O-}G(-P) \\ & + (A_{O1} + B_{O1}P)\big(F(RD_O - P) - F(-P)\big) + B_{O1}\big(G(RD_O - P) - G(-P)\big) \\ & + 2(A_{U2} + B_{U2}P)\big(F(CP - P) - F(RD_O - P)\big) + 2B_{U2}\big(G(CP - P) - G(RD_O - P)\big) \\ & + F(TG_O - P) - F(CP - P) \\ & + (A_{O2} + B_{O2}P)\big(F(1 - P) - F(TG_O - P)\big) + B_{O2}\big(G(1 - P) - G(TG_O - P)\big) \\ & + (A_{O+} + B_{O+}P)\big(1 - F(1 - P)\big) - B_{O+}G(1 - P)\end{aligned}$$

We know all the linear approximation coefficients in that expression, except $A_{O-}$, $A_{O+}$, $B_{O-}$, and $B_{O+}$. We solve for those by requiring that $E_O(0) = -L$ and $E_O(1) = W$, and that $A_{O-} = -L$ and $A_{O+} + B_{O+} = W$.

A similar expression can be found for the unavailable cube equity:

$$
\begin{aligned}
E_U(P) = {} & (A_{U-} + B_{U-}P)F(-P) + B_{U-}G(-P) \\
& + (A_{U1} + B_{U1}P)\big(F(TG_U - P) - F(-P)\big) + B_{U1}\big(G(TG_U - P) - G(-P)\big) \\
& - F(TG_U - P) + F(TP - P) \\
& + 2(A_{O1} + B_{O1}P)\big(F(RD_U - P) - F(TP - P)\big) + 2B_{O1}\big(G(RD_U - P) - G(TP - P)\big) \\
& + (A_{U2} + B_{U2}P)\big(F(1 - P) - F(RD_U - P)\big) + B_{U2}\big(G(1 - P) - G(RD_U - P)\big) \\
& + (A_{U+} + B_{U+}P)\big(1 - F(1 - P)\big) - B_{U+}G(1 - P)
\end{aligned}
$$

where $TG_U$ is the opponent's too-good point and $RD_U$ is the opponent's redouble point.

Finally for the centered-cube equity:

$$
\begin{aligned}
E_C(P) = {} & (A_{C-} + B_CP)F(-P) + B_{-C}G(-P) \\
& + (A_{C1} + B_{C1}P)\big(F(TGC_U - P) - F(-P)\big) + B_{C1}\big(G(TGC_U - P) - G(-P)\big) \\
& - F(TGC_U - P) + F(TP - P) \\
& + 2(A_{O1} + B_{O1}P)\big(F(ID_U - P) - F(TP - P)\big) + 2B_{O1}\big(G(ID_U - P) - G(TP - P)\big) \\
& + (A_{C2} + B_{C2}P)\big(F(ID_O - P) - F(ID_U - P)\big) + B_{C2}\big(G(ID_O - P) - G(ID_U - P)\big) \\
& + 2(A_{U2} + B_{U2}P)\big(F(CP - P) - F(ID_O - P)\big) + 2B_{U2}\big(G(CP - P) - G(ID_O - P)\big) \\
& + F(TGC_O - P) - F(CP - P) \\
& + (A_{C3} + B_{C3}P)\big(F(1 - P) - F(TGC_O - P)\big) + B_{C3}\big(G(1 - P) - G(TGC_O - P)\big) \\
& + (A_{U+} + B_{U+}P)\big(1 - F(1 - P)\big) - B_{U+}G(1 - P)
\end{aligned}
$$

where $TGC_U$ and $TGC_O$ are the too-good points for the opponent and player respectively, in the centered-cube case; and $ID_U$ and $ID_O$ are the initial double points for the opponent and player respectively.

To evaluate these expressions we need choices for $F$ and $G$. As noted earlier, and validated in Appendix 2, the choice of jump distribution does not much matter as long as the jump volatility is correct. I use the double-exponential distribution because it has simple expressions for $F$ and $G$.

The correct jump volatility to use in the $F$ and $G$ functions above is the local jump volatility when calculating equity at the current cube level, and the remote volatility when calculating equity at larger cube levels (for example, when estimating a redouble point). The linear approximation equities should use remote or local jump volatility as described earlier.

## Appendix 2: numerical confirmation of the small jump volatility approximation

Empirical jump volatilities are often around 10-15%, which is not as "small" compared to the market window as one might like. This section develops a numerical solution of the exact integral equations and compares it against the approximations developed in the previous appendix.

*Numerically estimating the cubeful equity*

The task before us is to solve the three integral equations for the cubeful equities developed in Appendix 1.

Discretize $P$=0% to 100% into $N$ buckets, where $P_0 = 0$ and $P_N = 1$. Consider just the first two integral equations for $E_O(P)$ and $E_U(P)$, as they depend only on each other and not $E_C(P)$. At the $i^{\text{th}}$ point,

$$E_{Oi} = \sum_{j=1}^{N} \int_{J=P_{j-1}-P_i}^{P_j-P_i} f_i(J)\hat{E}_O(P_i+J)dJ + I_{OLi} + I_{OWi}$$

$$I_{OLi} = \int_{J=-\infty}^{-P_i} f_i(J)E_O(P_i+J)dJ$$

$$I_{OWi} = \int_{J=1-P_i}^{\infty} f_i(J)E_O(P_i+J)dJ$$

I broke the integral over $J$ into $N$+2 pieces, and write the equity inside a bucket as $\hat{E}_O(P)$. Note that this isn't $E_O(P)$ - instead it is whatever equity is appropriate for that state, given doubling decisions. The last two terms correspond to the integrals for jumps below $P$=0 and above $P$=1. More on those in a moment.

Next assume that $\hat{E}_O(P)$ is piecewise linear, bounded by its value on either end of the bucket. This lets us write

$$E_{Oi} = \sum_{j=1}^{N} \left\{ \frac{\left((P_i-P_{j-1})\hat{E}_{Oj}+(P_j-P_i)\hat{E}_{Oj-1}\right)}{(P_j-P_{j-1})} \left(F(P_j-P_i)-F(P_{j-1}-P_i)\right) + \frac{(\hat{E}_{Oj}-\hat{E}_{Oj-1})}{(P_j-P_{j-1})} \left(G(P_j-P_i)-G(P_{j-1}-P_i)\right) \right\} + I_{OLi} + I_{OWi}$$

$\hat{E}_{Oj}$ is the post-jump equity at point j, which is determined by the optimal doubing decision at that point. $F$ and $G$ are as described in Appendix 1.

Now we can decide what to use for $\hat{E}_{Oj}$. This is determined by the optimal doubling decision. Here I make an additional assumption: we know the cube decision points already, and so know when to use the different alternatives for $\hat{E}_{Oj}$.

In fact, of course, we do not know the cube decision points. Instead, they should be the output of this solution, not the input. In practice we will iterate: guess values for the cube decision points, solve the system, then refine the cube decision points until convergence.

For the played-owned cube equation there are three relevant cube decision points: $RD_O$, the redouble point for the player; $CP$, the player's cash point; and $TG_O$, the player's too-good point. The post-jump equity $\hat{E}_{Oj}$ can be written in terms of those:

$$\hat{E}_{Oj} = \begin{cases} E_{Oj} & P < RD_O \text{ or } P > TG_O \\ 2E_{Uj} & RD_O < P < CP \\ 1 & CP < P < TG_O \end{cases}$$

That is, if *P* is below the player's redouble point or above his too-good point, he will choose not to double and stay at the current cube level. If *P* is between the redouble point and the cash point he will double and his opponent will take, so his equity jumps to the doubled unavailable cube equity. If *P* is above the cash point but below the too-good point, he will double but his opponent will pass, so he gets the cash amount.

Now we need to know what to do with the pieces of the integral for $P$<0 and $P$>1, $I_{OLi}$ and $I_{OWi}$. Really this should never matter – those pieces of the integral should always be zero. But to make that formally true we would need to use quite complex forms for the jump distribution, making the math even more complex.

One natural approach: assume that the equity for $P$<0 is always –L and for $P$>1 is always +W. That turns out not to be particularly appealing because practically that will lead to steps in equity across the $P$=0 and $P$=1 boundaries, even if we assume the jump volatility tends to zero at the boundaries.

A better approach is to assume that the equity is linear in those regions, extrapolating from the $i$=1 point for $P$<0 and from the $i$=$N$-1 point for $P$>1. The reason a linear extrapolation is nice is because a linear function solves those integral equations near the boundaries, so using a linear extrapolation ensures a nice linear function approaching $P$=0 and $P$=1. This gives us expressions for $I_{OL}$ and $I_{OW}$ of

$$I_{OLi} = \frac{E_{O1}}{P_1 - P_0}\left(G(-P_i) + P_i F(-P_i)\right) - \frac{L}{P_1 - P_0}\left(-G(-P_i) + (P_1 - P_i)F(-P_i)\right)$$

$$I_{OWi} = \frac{E_{O(N-1)}}{P_N - P_{N-1}}\left(G(1 - P_i) + (P_N - P_i)(1 - F(1 - P_i))\right) + \frac{W}{P_N - P_{N-1}}\left(-G(1 - P_i) + (P_i - P_{N-1})(1 - F(1 - P_i))\right)$$

We know $E_{O0} = -L$ and $E_{ON} = +W$, so only need to solve $E_{Oi}$ for $i$=1 to $N$-1. That gives us $N$-1 equations, but for potentially 2($N$-1) variables: $N$-1 for $E_{Oi}$ and $N$-1 for $E_{Ui}$, which also appears in the expressions.

We can repeat the process for the $N$-1 unavailable (opponent-owned) cube equities:

$$E_{Ui} = \sum_{j=1}^{N} \left\{ \frac{\left((P_i - P_{j-1})\hat{E}_{Uj} + (P_j - P_i)\hat{E}_{Uj-1}\right)}{(P_j - P_{j-1})} \left(F(P_j - P_i) - F(P_{j-1} - P_i)\right) + \frac{(\hat{E}_{Uj} - \hat{E}_{Uj-1})}{(P_j - P_{j-1})} \left(G(P_j - P_i) - G(P_{j-1} - P_i)\right) \right\} + I_{ULi} + I_{UWi}$$

with

$$\hat{E}_{Uj} = \begin{cases} E_{Uj} & P < TG_U \text{ or } P > RD_U \\ 2E_{Oj} & TP < P < RD_U \\ -1 & TG_U < P < TP \end{cases}$$

$$I_{ULi} = \frac{E_{U1}}{P_1 - P_0}(G(-P_i) + P_i F(-P_i)) - \frac{L}{P_1 - P_0}(-G(-P_i) + (P_1 - P_i)F(-P_i))$$

$$I_{UWi} = \frac{E_{U(N-1)}}{P_N - P_{N-1}}(G(1 - P_i) + (P_N - P_i)(1 - F(1 - P_i))) + \frac{W}{P_N - P_{N-1}}(-G(1 - P_i) + (P_i - P_{N-1})(1 - F(1 - P_i)))$$

I introduced three new cube decision points here: $TG_U$, the opponent's too-good point; $TP$, the take point; and $RD_U$, the opponent's redouble point. That brings us to six cube decision points. In order of $P$ they go: $TG_U$, $TP$, $RD_U$, $RD_O$, $CP$, and $TG_O$.

That gives us 2($N$-1) equations for our 2($N$-1) unknown equities. All the equations are linear in the equities, which means this is a simple linear system that we can solve with the usual linear algebra techniques.

Of course, this used our a priori guesses for the six cube decision points. We can then refine the cube decision points by looking at the solutions to the cubeful equities, solving:

$$E_U(TG_U) = -1$$
$$E_O(TP) = -\frac{1}{2}$$
$$E_U(RD_U) = 2E_O(RD_U)$$
$$E_O(RD_O) = 2E_U(RD_O)$$
$$E_U(CP) = \frac{1}{2}$$
$$E_O(TG_O) = 1$$

and making the assumption that the two different equities are piecewise linear in between the discrete points in *P*.

Once we solve for the new cube decision points we reconstruct the linear system and solve it again, then solve for the next iteration of cube decision points, repeating until the cube decision points no longer change from iteration to iteration. In practice this converges very quickly, in 2-4 iterations, with sensible initial guesses (for example, the cube decision points from the linear approximation).

This leads us to solutions for the player-owned and unavailable cube equities as well as the six cube decision points.

We then need to solve for the cube-centered equity, which is a function of the other two. In a similar process to before we can write

$$E_{Ci} = \sum_{j=1}^{N} \left\{ \frac{\left((P_i - P_{j-1})\hat{E}_{Cj} + (P_j - P_i)\hat{E}_{Cj-1}\right)}{(P_j - P_{j-1})} \left(F(P_j - P_i) - F(P_{j-1} - P_i)\right) + \frac{(\hat{E}_{Cj} - \hat{E}_{Cj-1})}{(P_j - P_{j-1})} \left(G(P_j - P_i) - G(P_{j-1} - P_i)\right) \right\} + I_{CLi} + I_{CWi}$$

with

$$\hat{E}_{Cj} = \begin{cases} E_{Cj} & P < TG_{UC} \\ -1 & TG_{UC} < P < TP \\ 2E_{Oj} & TP < P < ID_U \\ E_{Cj} & ID_U < P < ID_O \\ 2E_{Uj} & ID_O < P < CP \\ 1 & CP < P < TG_{OC} \\ E_{Cj} & P > TG_{OC} \end{cases}$$

$$I_{CLi} = \frac{E_{C1}}{P_1 - P_0}(G(-P_i) + P_i F(-P_i)) - \frac{L}{P_1 - P_0}(-G(-P_i) + (P_1 - P_i)F(-P_i))$$

$$I_{CWi} = \frac{E_{C(N-1)}}{P_N - P_{N-1}}(G(1 - P_i) + (P_N - P_i)(1 - F(1 - P_i))) + \frac{W}{P_N - P_{N-1}}(-G(1 - P_i) + (P_i - P_{N-1})(1 - F(1 - P_i)))$$

As before we have $E_{C0} = -L$ and $E_{CN} = +W$, so there are *N*-1 equities to solve, and we have *N*-1 equations.

There are four new cube decision points, however: $TG_{UC}$, the too-good point for the opponent when the cube is centered; $ID_U$, the opponent's initial double point; $ID_O$, the player's initial double point; and $TG_{OC}$, the player's too-good point. As before we make an initial guess at these, solve the resulting linear system, recalculate the cube decision points, and iterate until convergence.

This outputs the $N$-1 values for the cube-centered equity as well as the four new cube decision points. The final result is the $N$-1 values for all three equities plus ten cube decision points.

To implement this solution we need to choose a distributional form for the jump distribution. Two simple choices are Gaussian and double-exponential. Both are symmetric distributions; the Gaussian distribution has a kurtosis of zero and the double-exponential a kurtosis of 3, so they give different tail behavior. Some details of the two distributions are given below:

| | Gaussian | Double-Exponential |
|---|---|---|
| $f(J)$ | $\frac{e^{-\frac{J^2}{2\sigma^2}}}{\sigma\sqrt{2\pi}}$ | $\frac{\lambda}{2}\begin{cases} e^{-\lambda J} & J > 0 \\ e^{+\lambda J} & J \leq 0 \end{cases}$ |
| $F(J)$ | $N\left(\frac{J}{\sigma}\right)$ | $\begin{cases} 1 - \frac{1}{2}e^{-\lambda J} & J > 0 \\ \frac{1}{2}e^{+\lambda J} & J \leq 0 \end{cases}$ |
| $G(J)$ | $-\frac{\sigma}{\sqrt{2\pi}}e^{-\frac{J^2}{2\sigma^2}}$ | $\begin{cases} -\frac{1}{2}\left(\frac{1}{\lambda}+J\right)e^{-\lambda J} & J > 0 \\ -\frac{1}{2}\left(\frac{1}{\lambda}-J\right)e^{+\lambda J} & J \leq 0 \end{cases}$ |
| Standard deviation | $\sigma$ | $\frac{\sqrt{2}}{\lambda}$ |
| Jump volatility $\alpha = E[\lvert J \rvert]$ | $\sigma\sqrt{\frac{2}{\pi}}$ | $\frac{1}{\lambda}$ |

where $N(x)$ is the cumulative standard normal distribution function.

An important result of the numerical experiments detailed below is that the cubeful equity calculations depend on the jump volatility of the jump distribution, but not any other features of the distribution. This is a convenient result, as we need only to track the jump volatility and estimate how it changes with game state, not also other features such as standard deviation or kurtosis.

Since the distribution features are not important I will tend to use the double-exponential distribution for jumps. For this distribution $F$ and $G$ do not rely on any special functions (such as the standard cumulative normal distribution function).

The next figure shows a representative solution using the double-exponential distribution, with $W$=1.2, $L$=1.1, and jump volatility $\alpha$=8%. The equity as a function of cubeless win probability is roughly linear in most regions but curves around the cash point (player-owned and centered) and take point (opponent-owned and centered).

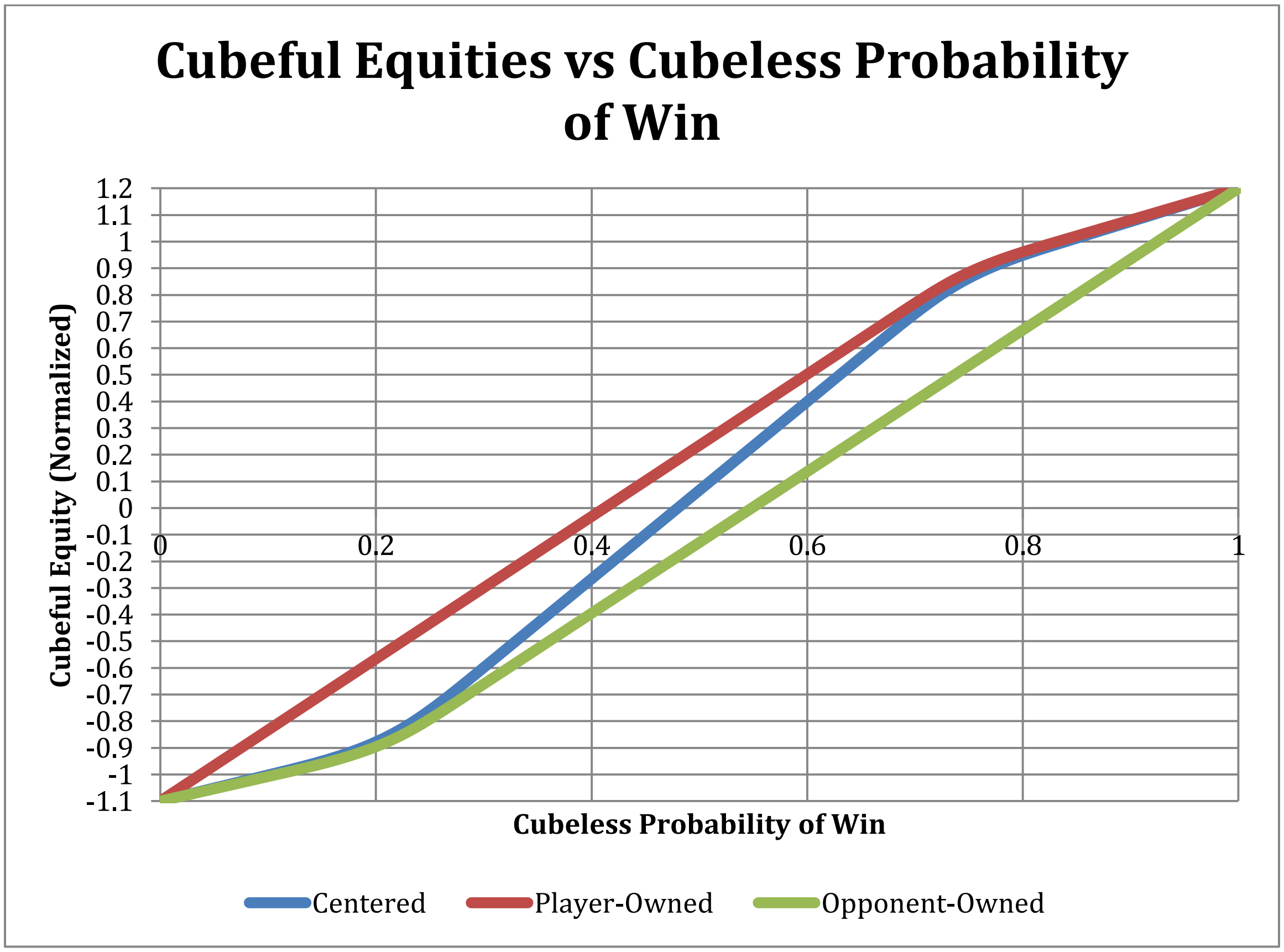

Figure 3: cubeful equities vs probability of win for the three cube states (centered, player-owned, and opponent-owned). All equities are normalized by the cube value, so all of them equal –L at P=0 and +W at P=1. The jump distribution was double-exponential with jump volatility $\alpha$=8%, W=1.2, and L=1.1.

For that same game structure we can compare the results using a Gaussian jump distribution and a double-exponential distribution with the same jump volatility. Note that I am keeping the jump volatility the same between the distributions, not the standard deviation, because as we saw in the development of the approximation, the jump volatility is the more relevant quantity.

The next figure shows a comparison of the cubeful equity under the two models for the centered-cube case. The resulting cubeful equities are very close: the equity mainly depends on jump volatility, not other features of the jump distribution.

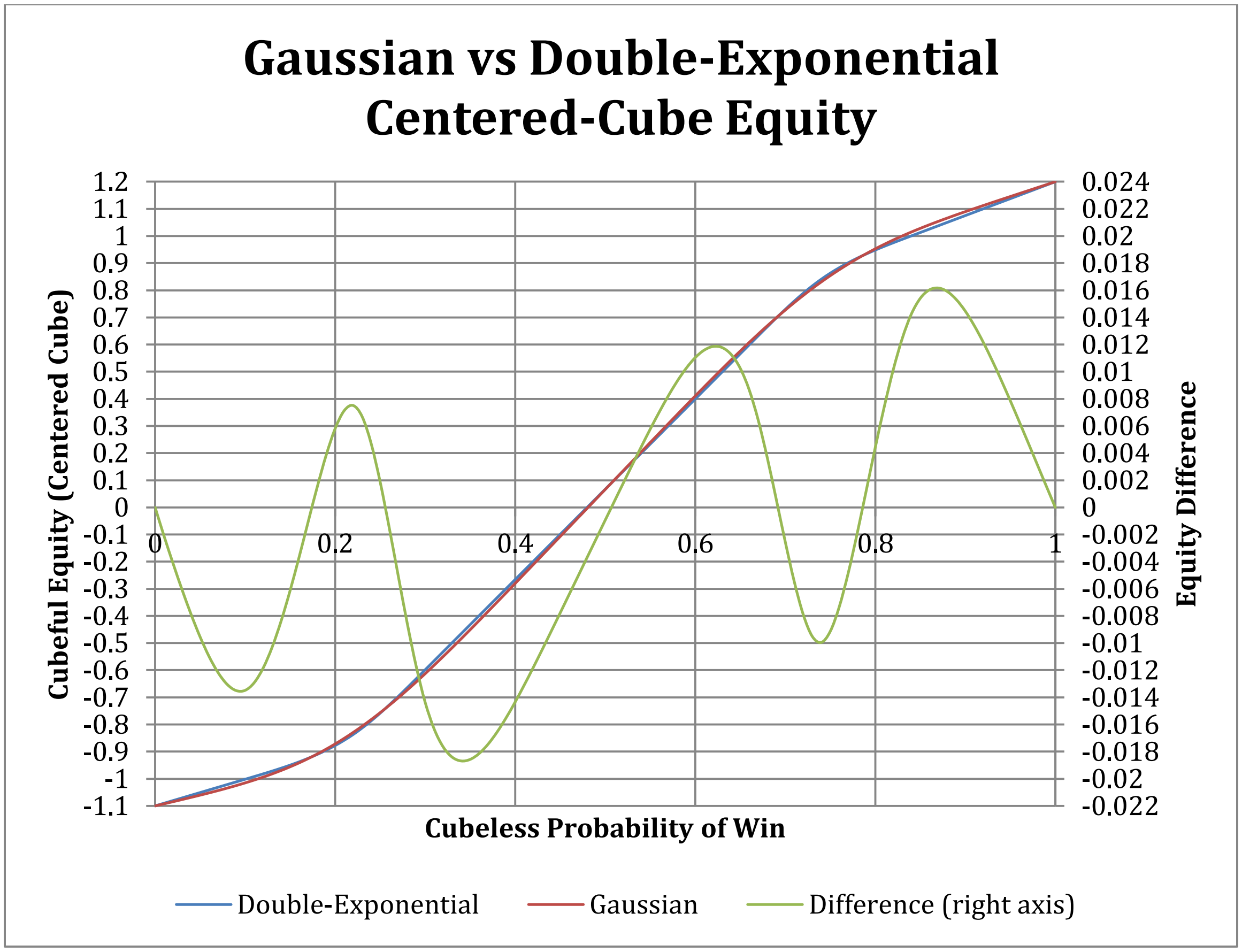

Figure 4: comparison of cubeful equity (with a centered cube) for a Gaussian jump distribution and a double-exponential jump distribution, both with the same jump volatility. W=1.2, L=1.1, $\alpha$=8%. The blue and red lines show the model equities on the left axis; the green line shows the difference, with an expanded range on the right axis. Equity differences are small if the same jump volatility is used; the results are otherwise relatively insensitive to the form of the jump distribution.

Next I compare the numerical calculation with the approximation results, both the linear approximation and the nonlinear one. I chose $W$=1.4, $L$=1, and $\alpha$=20% as a relatively extreme and asymmetric case to stress the approximation.

The linear approximation does a reasonable job of matching the exact calculation, especially away from the take and cash points. The nonlinear approximation does better, especially in the centered-cube case, and results in more accurate cube decision point estimates.

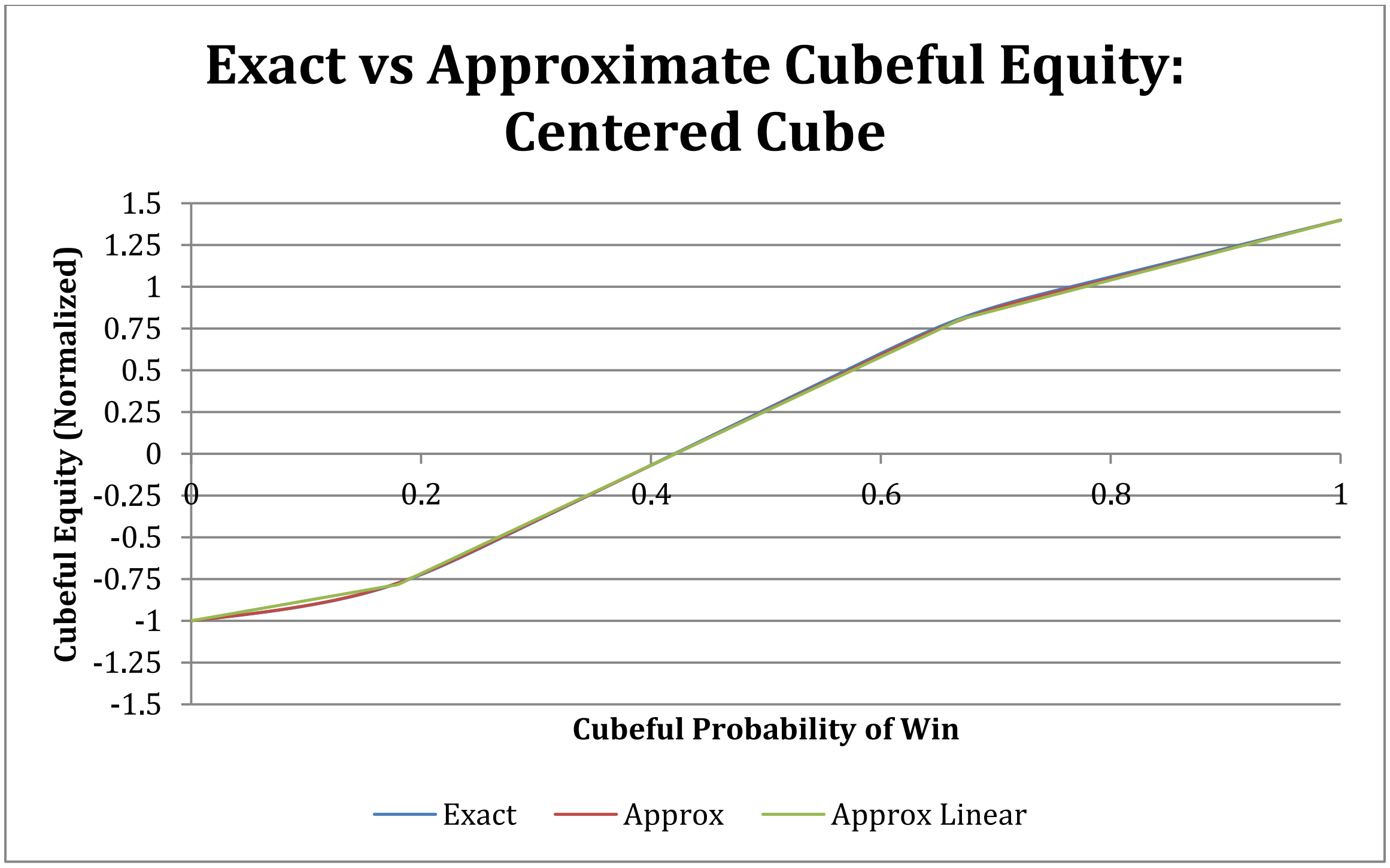

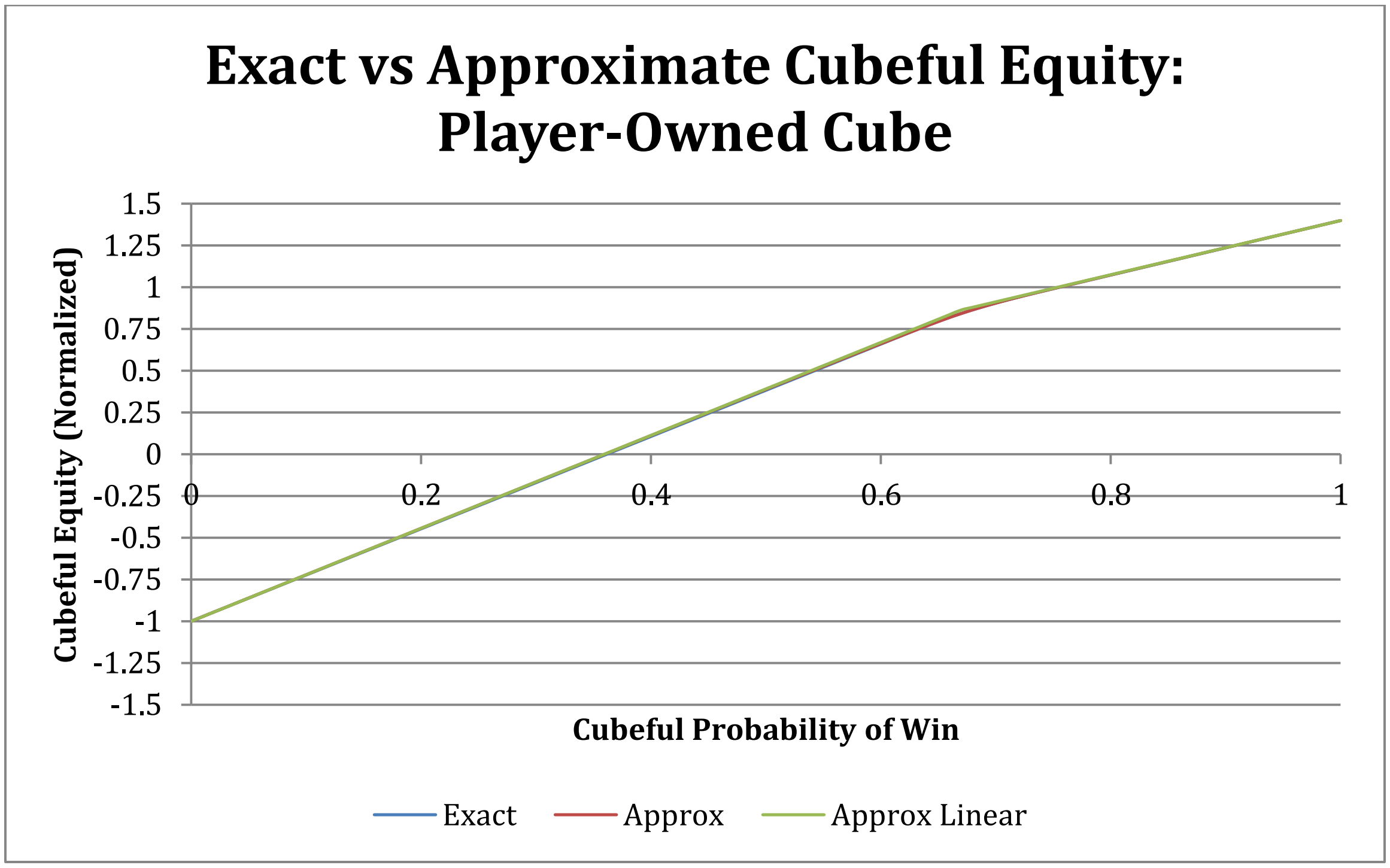

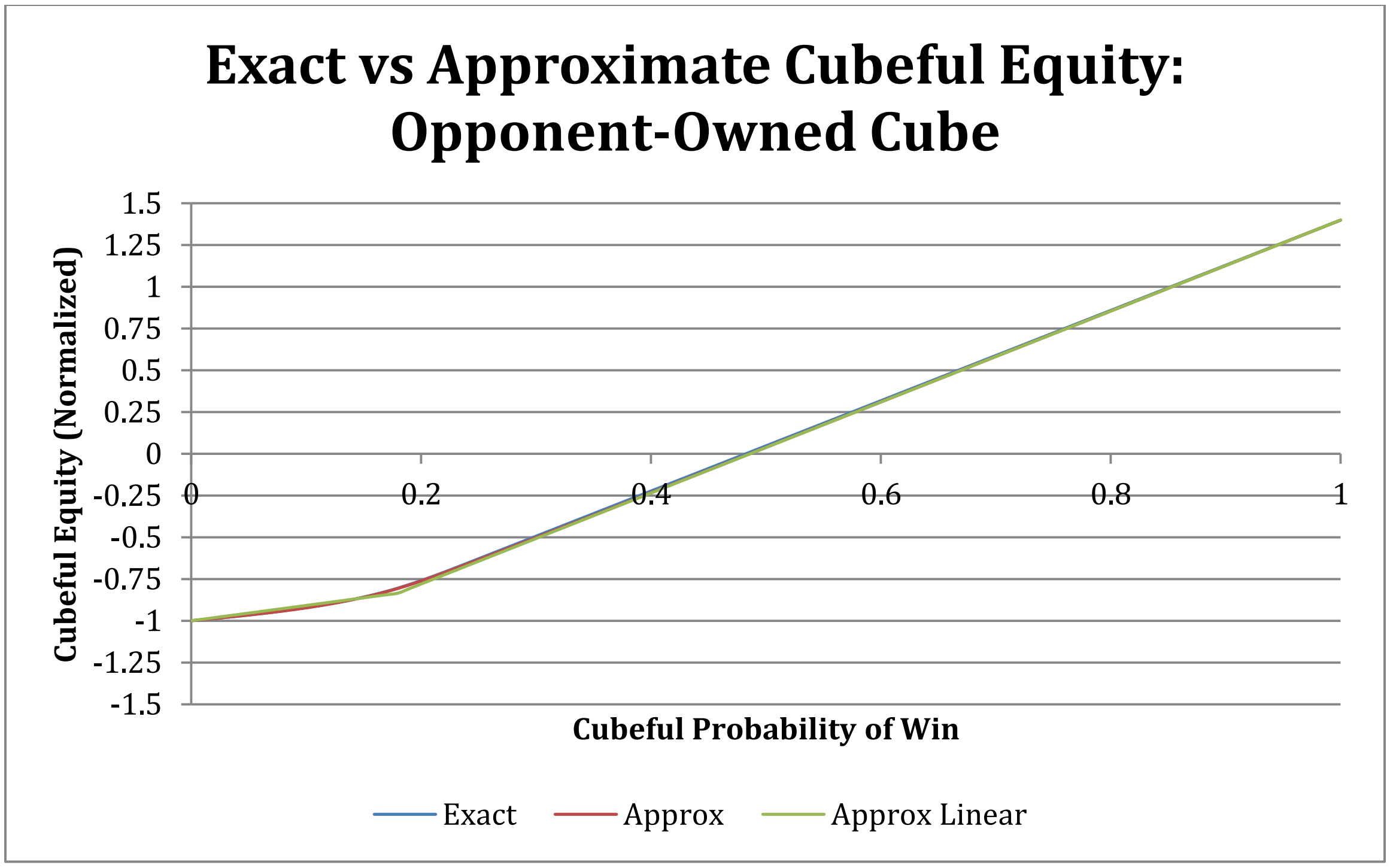

Figures 5-7: cubeful equity for the centered-cube, player-owned cube, and opponent-owned cube cases (normalized by the cube value). The blue line shows the exact numerical calculation of equity, the red line shows the nonlinear approximation, and the green line shows the linear approximation. $W$=1.4, $L$=1, and $\alpha$=20%, an extreme scenario meant to stress the approximation. The linear approximation is quite close, but the nonlinear approximation is much closer.

# References

1. Janowski, R. "Take-Points in Money Games", bkgm.com, http://www.bkgm.com/articles/Janowski/cubeformulae.pdf
2. Keeler, E and Spencer, J. "Optional Doubling in Backgammon", Operations Research, Vol. 23, No. 6, November-December 1975.
3. Johansen, Ø. Post about Janowski's formula in rec.games.backgammon Usenet forum. http://groups.google.com/group/rec.games.backgammon/msg/06f9726a0e116fd7
4. GNU Backgammon online documentation. http://www.gnubg.org/documentation/doku.php?id=appendix#12
5. Higgins, M. Computational Backgammon blog, http://compgammon.blogspot.com